# A Heuristic Method for Load Retrievals Route Programming in Puzzle-based Storage Systems

Yang Zou [1], Mingyao Qi [1*]

[1] *Graduate School at Shenzhen, Tsinghua University, Shenzhen 518055, China*

**Abstract:**

Recently, many enterprises are facing the difficulties brought out by the limitation of warehouse land and the increase of loan cost. As a promising approach to improve space utilization rate, puzzle-based storage systems (PBSSs) are drawing more attention from logistics researchers. In previous research about PBSS, concentration has been paid to single target item problems. However, there are no consensus algorithms to solve load retrievals route programming in PBSSs with multiple target items. In this paper, a heuristic algorithm is proposed to solve load retrievals route programming in PBSSs with multiple target items, multiple escorts and multiple IOs. In this paper, new concepts about the proposed algorithm are defined, including target IOs, target position of escorts, number of escorts required, et al. Then, the decision procedures are designed according to these concepts. Based on Markov decision process, the proposed algorithm makes the decision of the next action by analyzing the current status, until all the target items arrive at the IOs. The case study results have shown the effectiveness and efficiency of the proposed heuristic algorithm.

1. **Introduction**

Recently, with the development of the economy of the society, smart logistics has drawn more attention. As a result, logistics industry is developing towards a kind of automated, unmanned, and smart mode. Smart storage is a vital part of smart logistics. Many enterprises are facing the difficulties brought out by the limitation of warehouse land and the increase of loan cost. In some cities, the price of industrial land has shown a rapid upward trend. In 2012, the price of industrial land in Guangzhou increased by more than 20%. Under the background of recent economic globalization, the market competitiveness of enterprises has been reduced, and tremendous pressure has been put on the development of enterprises. Therefore, how to make full use of land resources and maximize storage efficiency under limited land resources has become a practical and challenging research field in smart logistics.

___________

*Corresponding author. Email address: qimy@sz.tsinghua.edu.cn

As a promising approach to improve space utilization rate, dense storage is drawing more attention from logistics researchers. In traditional warehouses, the lanes between shelves provide transportation space for the storage and retrieval of goods. However, the lanes themselves cannot be applied to store goods. The existence of lanes has led to a reduction in the space utilization of the storage system. In order to improve the space utilization of the warehouse, some scholars have proposed a puzzle-like dense storage, called the puzzle-based storage systems (PBSSs). In a PBSS, objects are stored in each storage unit. There is no fixed lane in a PBSS, but only the positions of the escorts where the goods are not stored. Each storage unit has a movement unit to control its movement direction, so that it can move autonomously. Then, warehouse users can set the number of escorts in the system according to practical needs. Compared with the traditional fixed aisle warehouse storage method, the PBSS can greatly improve the utilization of space resources and store as many goods as possible in a limited space. In addition, compared with traditional dense storage models such as automated storage and retrieval system (AS&RS), the PBSS can set up the warehouse sites reasonably and flexibly, thereby reducing the cost of using the warehouse and meeting the needs of enterprises for warehousing land.

However, the weakness of PBSS is that it is more difficult to store and retrieve goods, because that there is no lane in PBSS as in traditional warehouses. As a result, the picking of goods can no longer be completed by a simple grasping action. It is necessary to move other goods that block the target goods through a series of continuous actions, in order to retrieve the target goods from the IOs. By programming the movement route of the goods, the purpose of retrieving the goods can be achieved with less movement, and the efficiency of storing and retrieving goods can be improved. Therefore, the route programming of load retrieval for PBSS has huge practical application value.

In previous literatures, the load retrieval route programming problem for a PBSS with only one target item has been considered. As for the PBSS with multi-target items, it has rarely been touched in previous research. This paper innovatively proposes a heuristic algorithm for solving the load retrieval route programming problem of a PBSS with multi-target items, multiple escorts, and multi-IOs, which complements the research in this field. The algorithm proposed in this paper can be applied to solve problems in practical industrial production, which has strong application value.

The rest of this paper is organized as follows. The literature review is given in Section 2. In Section 3, the background about PBSS is introduced. In Section 4, the heuristic algorithm to solve the load retrieval routing problem in PBSSs is proposed. Numerical results for different cases are given and discussed in Section 5. Section 6 concludes this paper.

## 2. Literature Review

In a PBSS, goods are stored on movable storage units, and items and escorts are randomly distributed in the system. The route programming problem of PBSS load retrieval is to plan the moving routes of the target items, so as to move the target items to the IOs with the least number of moving steps. Scholars have conducted research on the optimization of load retrieval routes in PBSSs. According to the number of target items in a PBSS, it can be divided into two categories: single-target-item and multi-target-item PBSS problems. In addition, the number of escorts and IOs can also be one or multiple. TABLE I summarizes the state of the art of PBSS load retrieval route programming problems.

TABLE I Literatures about PBSS

| Approach | Year | Method | Number of Target Items | Number of Escorts | Number of IOs |
|---|---|---|---|---|---|
| [1] | 2007 | Single escort: exact algorithm / Multiple escorts: heuristic algorithm | 1 | Multiple | 1 |
| [2] | 2010 | Heuristic | 1 | Multiple | 1 |
| [3] | 2015 | ≤2 target items: exact algorithm / >2 target items: heuristic algorithm | 1 | Multiple | 1 |
| [4] | 2018 | Small scale: exact algorithm / Large scale: heuristic algorithm | 1 | Multiple | Multiple |
| [5] | 2014 | Distributed control | Multiple | Multiple | Multiple |
| [6] | 2017 | ≤2 target items: exact algorithm / >2 target items: heuristic algorithm | Multiple | 1 | 1 |
| **This paper** | **2021** | **Heuristic algorithm** | **Multiple** | **Multiple** | **Multiple** |

For the PBSS with one target item, there have been many related research results. Gue proposed the concept of PBSS in 2007 [1], in which he pointed out that this dense storage method is most likely to store goods with the highest density. For a PBSS with only one target item, one escort, and one IO, Gue et al. established an integer programming mathematical model and gave an analytical solution to the optimal route. For a specific PBSS with one target item, multiple escorts, one IO, and the escorts are arranged in a line next to the IO, Gue et al. proposed a heuristic algorithm based on map partition to solve this type of problem. By comparing with the operating efficiency of the tunnel-type non-intensive storage mode, it is concluded that the storage density of the PBSS is higher, but the access speed is lower. In practical applications, it should be selected according to specific goals and constraints. In 2010, Alfieri et al. proposed a heuristic strategy to solve the single-target item, multi-escort PBSS based on the exact algorithm proposed by Gue for the single-target item and single-escort problem [2]. They have added the limit that the number of movable units is less than the number of stored items. In this type of warehouse, the storage unit under each item cannot move autonomously and requires the assistance of an automated guided vehicle (AGV) to move. When the AGV is not under the storage unit, the cargo storage unit cannot move autonomously. They first considered the balance of AGV tasks in the algorithm, assigned tasks to existing AGVs, and then designed management strategies for the movement paths of these AGV-assisted storage units.

In 2015, Kota et al. studied the number of moving steps required to move a single target item from its initial position to an IO [3]. For the situation that the number of escorts is no larger than 2, the authors proposed an analytical solution expression and demonstrated that the number of moving steps is largely affected by the distribution of escorts. For the case where the number of escorts is larger than 2, the authors propose a heuristic algorithm. The biggest difference between this paper and the previous research is that the previous research only considered a certain specific arrangement of escorts, while this research considered the different distribution of escorts in PBSS. This is more instructive for practical applications. In 2018, with the goal of minimizing the number of moving steps, Altan et al. studied the route programming problem with one target item, multiple escorts, and multiple IOs in PBSS [4]. The researchers abstracted the problem as a search problem in the state space. Based on the A* algorithm, they proposed a search algorithm that can obtain the optimal solution and then pruned the algorithm. For large-scale cases, the authors propose a heuristic modification of the search algorithm, and proves the quasi-optimality of the solutions and the effectiveness of the algorithm through experiments.

For the PBSS path optimization problem of multi-target items, the current research results are very limited due to the higher complexity of the problem. In 2014, Gue et al. proposed a PBS control system based on decentralized control, and each module runs independently according to the designed operating logic [5]. The author proposes a simpler motion rule to avoid deadlock. However, this system requires that the target item can only enter from one side of a PBSS and exit from the opposite side, which has many limitations in practical applications. Mirzaei et al. studied the PBSS path optimization algorithm with multi-target items, single escort, and single IO [6]. For the situation where there are two target items, the author first let the two target items merge, and then advance to the IO port together. By assuming that different items can move at the same time, the authors propose an exact algorithm based on the integer programming model, and compares the results of the heuristic algorithm with the exact algorithm. For the case where the number of target items is greater than two, the author designs a heuristic algorithm based on this idea to solve the problem.

## 3. Puzzle-based Storage System and its Route programming

In this section, the elements and status transition in a PBSS and the problem formation of PBS route programming are introduced and described.

### 3.1. Elements in a PBSS

The elements in a PBSS include the target items, other items, the escorts and the IOs. A PBSS is usually described by a map. The elements are introduced as follows:

Target item: the items which are to be retrieved from the PBSS. The aim of the PBS route optimization algorithm is just to retrieve the target items with as few as possible moving actions.

Other item: the items which are also stored in the PBSS but are not to be picked up by the route optimization algorithm. However, these items occupy the positions in the PBSS and are regarded as the obstacles in retrieving the target items.

Escort: the positions in the map which are not occupied by any item. By exchanging the positions of an escort and an item, the item can be moved. In this process, the number of escorts in a PBSS remains unchanged.

IO: the pre-set exit positions for items in a PBSS. Any items or escorts can be placed at the IOs during operation. When a target item arrived at an IO, the retrieval of it is achieved.

Fig. 1 shows an example 5*5 PBSS with 3 target items, 4 escorts and 3 IOs. In this paper, the black squares in a map refer to the target items, the white squares refer to the escorts, the grey squares refer to other items, and the red boxes refer to the positions of the IOs. In Fig. 1, the axis of 3 IOs are (0, 4), (2, 4) and (4, 4), respectively.

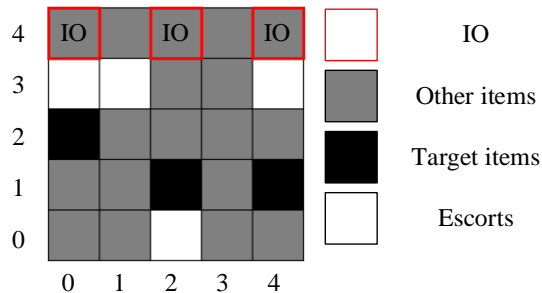

Fig. 1 A 5×5 PBSS with 3 target items, 4 escorts and 3 IOs.

### 3.2. Status Transition in a PBSS

In this paper, the item retrieval problem of a PBSS is regarded as a discrete Markov decision process (DMDP). In a DMDP, the transition of status is accomplished by actions. Therefore, the status and action in the item retrieval problem of a PBSS should be defined first. The status $s$ in this DMDP refers to the arrangement of the target items, other items and the escorts in a PBSS. The action $a$ in this DMDP is to move an escort in a status in order to transit into another status $s'$. From one original status $s$, multiple potential neighbour statuses $s_i'$ can be entered by different actions $a_i$, forming a set of neighbour statuses $\{s_i'\}_s$.

An example of the status transition in a 4*4 PBSS is given in Fig. 2. The current status is s, all the possible actions include $a_1$, $a_2$, $a_3$ and $a_4$, and the neighbour statuses are $s_1'$, $s_2'$, $s_3'$ and $s_4'$.

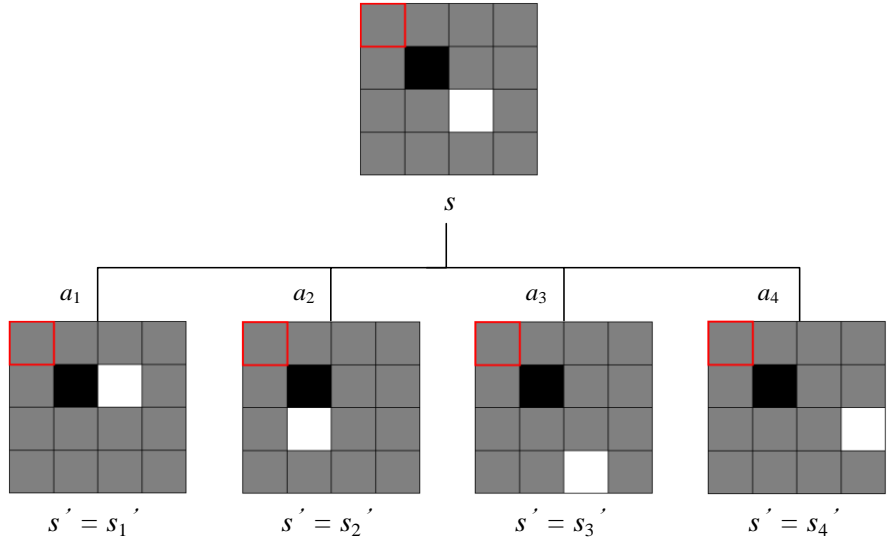

Fig. 2 An example of PBSS status transition

### 3.3. PBSS Route Optimization

PBSS route optimization is to program the moving routes of the escorts and the target items, so that the target items can be moved to the IOs with minimal actions for a given PBS. In this paper, the route optimization of a PBSS with multiple target items is considered based on the previous research about the route optimization of a PBSS with single target item.

As introduced previously, the moving of a target item is achieved by the moving of an escort which is adjacent to the target item. An example of the moving of a target item is given in Fig. 3. To move the target item from its original position (1, 0) to the IO (0, 0), the escort A is firstly moved from its original position (0, 2) to the target position of the target item (0, 0), as shown in Fig. 3 (1), (2) and (3). Afterwards, the positions of the escort and the target item are exchanged, as shown in Fig. 3 (3) and (4). In this way, the target item is moved to its target position.

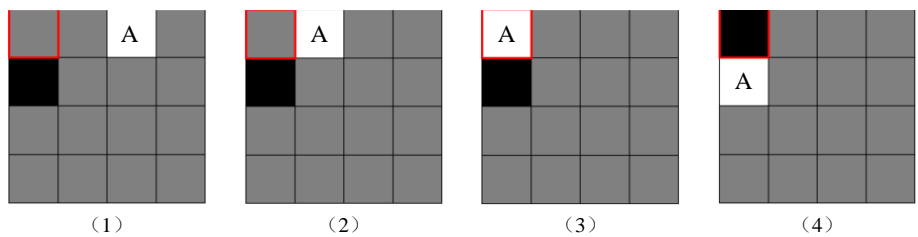

(1)　　　　　　　(2)　　　　　　　(3)　　　　　　　(4)

Fig. 3 An example of the moving of a target item

When the PBS route optimization problem is expanded from single target item to multiple target item, it is more complicated for the following reasons. Firstly, the routes of multiple target items may have 交互 impacts among each other. In contrast, in the PBS route optimization problem for single target item, only the moving of one target item towards the IO needs to be considered. Secondly, as a result of the first aspects, the coordinated application of one escort in the routes of multiple target items should be considered. After moving one target item, it would be better to move the escort to a position which makes it applicable for moving other target items. In this way, the total moving steps can be saved. Thirdly, as there are multiple target items and IOs in a PBSS, the distribution of the IOs for the target items should be considered, i.e. which target item should be moved towards which IO. This will be further discussed in the algorithm section.

4. **A Heuristic Algorithm for Route programming**

In this section, a heuristic algorithm for load retrieval in a PBSS is proposed. Load retrieval in a PBSS is a sequential decision making problem in a DMDP. In the proposed heuristic algorithm, the sequential decisions are made based on the current PBSS status in order to retrieve all the target items towards the IOs with minimal moving steps.

The heuristic algorithm makes the sequential decisions after the evaluation and analysis of the current status, which mainly include four steps: the optimal distribution of target IOs, the determination of escort target positions, the computation of the number of required escorts for the target items, and the computation of the distance matrix **D**. After these four steps, the current status is analyzed in detail and the next moving decision is made based on the following metrics: the minimal total distance between the target items and the IOs ($d_{s\_min}$), the minimal total number of required escorts for the target items ($et_{s\_min}$), and the minimal item of the distance matrix **D** ($\min(\mathbf{D}_s)$). Then, the decision of the next action can be made based on these three metrics.

*4.1. Optimal Distribution of Target IOs*

The distribution of target IOs is to determine which IO should one target item be moved towards. This is a new problem in the load retrieval problems of PBSSs with multiple target items and multiple IOs. In other words, there is no necessity to consider this problem in the load retrieval problems with single target item and single IO. The basic rule to distribute target IOs for the target items is to minimize the total distance between the target items and their corresponding target IOs, which is denoted as *d*.

Assume that in a PBSS map there are $l$ IOs, which are located at A={$\mathbf{io}_1$, $\mathbf{io}_2$, …, $\mathbf{io}_l$}, $\mathbf{io}_i \in \mathbf{R}^{1\times 2}$, $i=1, 2, …, l$, and $l$ target items, which are located at B={$\mathbf{m}_1$, $\mathbf{m}_2$, …, $\mathbf{m}_l$}, $\mathbf{m}_j \in \mathbf{R}^{1\times 2}$, $j=1, 2, …, l$. The Manhattan distance between $\mathbf{m}_j$ and $\mathbf{io}_i$ is defined as $e_{i,j}$. For a permutation of $\{1, 2, …, l\}$, $[k_1, k_2, …, k_l]$, a distribution plan of the target IOs is to distribute the target item M$_{kj}$ which is located at **m**$_j$ to IO$_{kj}$ whose location is **io**$_{kj}$, $j=1, 2, …, l$. In this way, a one-on-one mapping between the positions of the IOs A and those of the target items B is formed. For this distribution plan, the total distance between $l$ target items and their corresponding target IOs is:

$$d(k_1, k_2, ..., k_l) = \sum_{j=1}^{l} e_{kj,j} \quad (1)$$

where $e_{kj,j} = |x_{io_{kj}} - x_{m_j}| + |y_{io_{kj}} - y_{m_j}|$, $(x_{io_{kj}}, y_{io_{kj}})$ and $(x_{m_{kj}}, y_{m_{kj}})$ are the coordination of **io**$_{kj}$ and **m**$_j$, respectively. The optimal distribution of target IOs is to find $h$ permutations $[o_1, o_2, …, o_l]_i$, $i=1, 2, …, h$, among all the $A^l$ permutations of $\{1, 2, …, l\}$, in order to minimize $d$ to its minimal value $d_{s\_min}$. For a PBSS status, there may be multiple distribution plans of target IOs which can minimize $d$. Then, all these distribution plans will be regarded as the optimal distribution plans. In other words, a target item may have multiple target IOs. This will be further illustrated in the case study section.

*4.2. Determination of Escort Target Positions*

To begin with, the definition of escort target positions is given. A position $p$ is regarded as an escort target position if and only if it satisfies the following two conditions:

(1) It is adjacent to a target item, i.e.:

$$|x_1 - x| + |y_1 - y| = 1 \quad (2)$$

where $(x_1, y_1)$ is the coordination of a target item and $(x, y)$ is the coordination of the position $p$.

(2) If the adjacent target item is moved to $p$, the distance between the target item and one or multiple target IOs can be shorten, i.e.:

$$\exists i, s.t. (x_i - x_1)(x - x_1) > 0 \text{ or } (y_i - y_1)(y - y_1) > 0, i = 1, 2, ...h$$

where $(x_i, y_i)$ is the coordination of a target IO.

In this way, all the positions which satisfy the above two conditions form the set of escort target positions. If an escort is moved to a escort target position, it can be further applied to move the adjacent target item towards one or multiple target IOs, after which $d$ can be reduced.

*4.3. Computation of the Number of Required Escorts for Target Items*

The number of required escorts for a target item is an important index which will be applied in the next subsection: the computation of distance matrix **D**. Firstly, the definition is given as follows. For a target item M$_j$, whose target IO is IO$_{oj}$ with the distribution of target IOs as $[o_1, o_2, …, o_l]_i$, the number of required escorts in the route between M$_j$ and IO$_{oj}$ is defined as follows:

$$en_{ij} = e_{oj,j} - es_{oj,j} = |x_{IO_{oj}} - x_{M_j}| + |y_{IO_{oj}} - y_{M_j}| - es_{oj,j} \quad (3)$$

where $e_{oj,j}$ is the relative distance between $M_j$ and $IO_{oj}$, and $es_{oj,j}$ is the number of existing escorts in the route between $M_j$ and $IO_{oj}$, which is defined as the number of escorts within the rectangle formed by $M_j$ and $IO_{oj}$. An example is given in Fig. 4. The distance between $M_1$ and $IO_1$ is 3. Within the rectangle formed by $M_1$ and $IO_1$, the number of exiting escorts is 1. As a result, the number of required escorts for $M_1$ is 2.

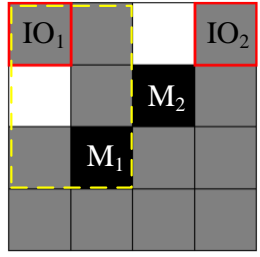

Fig. 4 The number of required escorts from $M_1$ to $IO_1$.

Based on the definition of $en_{ij}$, the total number of required escorts for all the target items under a given distribution plan of the IOs $[o_1, o_2, …, o_l]_i$ can be calculated as follows, where $s$ denotes the current status.

$$et_{si} = \sum_{j=1}^{l} en_{ij}, \forall i \in \{1, 2, ..., h\} \quad (4)$$

Further, for all the $h$ distribution plans, the corresponding $et_{si}$ for $i=1, 2, …, h$ can be calculated. Then, the minimal value is denoted as $et_{s\_min}$, which will be applied in further sections.

*4.4. Computation of Distance Matrix $\mathbf{D}_s$*

4.4.1. Structure of Distance Matrix $\mathbf{D}_s$

The distance matrix $\mathbf{D}_s$ is to measure the required steps of moving the escorts to the escort target positions, which is an important status evaluation index in the proposed heuristic algorithm. Assume that in status $s$ the positions of escorts are $\{\mathbf{ep}_1, \mathbf{ep}_2, …, \mathbf{ep}_n\}$, $\mathbf{ep}_i \in \mathbf{R}^{1\times 2}$, $i=1, 2, …, n$, and the positions of escort target positions are $\{\mathbf{t}_1, \mathbf{t}_2, …, \mathbf{t}_m\}$, $\mathbf{tp}_j \in \mathbf{R}^{1\times 2}$, $j=1, 2, …, m$. Then, the structure of $\mathbf{D}_s$ is given as follows:

$$\mathbf{D}_s = \begin{bmatrix} c_{1,1} & c_{1,2} & \cdots & c_{1,m} \\ c_{2,1} & c_{2,2} & \cdots & c_{2,m} \\ \vdots & \vdots & \ddots & \vdots \\ c_{n,1} & c_{n,2} & \cdots & c_{n,m} \end{bmatrix}$$

where $c_{i,j}$ is the estimated moving steps from $\mathbf{ep}_i$ to $\mathbf{tp}_j$. Then, all the items in the same row represent the estimated moving steps from one escort to all the escort target positions. Also, all the items in the same column represent the estimated moving steps from all the escorts to one escort target position.

4.4.2. Distance between an Escort and an Escort Target Position

In this section, the method to compute $c$ is proposed. Basically, the distance between an escort E ($x_E$, $y_E$) to an escort target position T ($x_T$, $y_T$) is:

$$c = e_{E,T} + t + r = |x_E - x_T| + |y_E - y_T| + t + r \tag{5}$$

where $t$ and $r$ are two types of corrections, which will be introduced as follows. They are both introduced in order to better estimate the moving steps from an escort to an escort target position and then to provide better quantification basis for decision making.

(1) Correction $t$

Correction $t$ is introduced to deal with the cases in which some target items are blocking the direct path between the escort and the escort target position. Three examples are given in Fig. 5. In example (a), a target item is blocking the direct path (Path 1) between E and T. As a result, E has to be moved through Path 2, in which two additional steps are needed. In example (b), two target items block all the direct paths (Path 1, 2, 3) between E and T, and E has to be moved through Path 4, in which two additional steps are needed. In both cases, $t=2$.

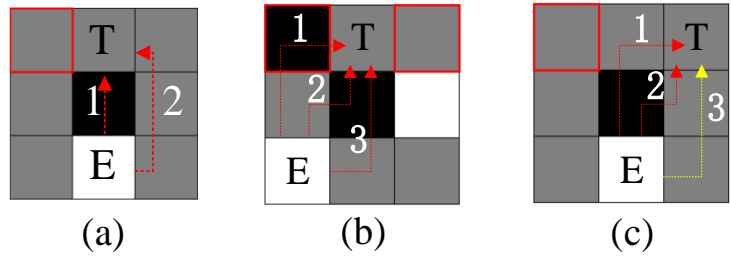

Fig. 5 Three examples to illustrate correction $t$.

The method to compute $t$ is to list all the direct paths between E and T, after which whether all the paths are blocked can be judged. The total number of direct paths is $C_{|x_E-x_T|+|y_E-y_T|}^{|x_E-x_T|}$. If all the direct paths are blocked, such as the example cases (a) and (b), $t=2$. In contrary, if there exists one path which is not blocked, $t=0$. An example case of $t=0$ is given in Fig. 5 (c). In this case, although Path 1 and Path 2 are blocked by a target item, Path 3 is clear and E can be moved through this path to reach T. Therefore, in this case $t=0$.

(2) Correction $r$

The aim of the introduction of correction $r$ is to estimate the moving steps more accurately, in which the reuse of escorts is considered. When moving a target item M ($x_M$, $y_M$) to its target IO ($x_{IO}$, $y_{IO}$) through an escort target position T ($x_T$, $y_T$), a rectangle is formed by two points ($x_T$, $y_T$) and ($x_{IO}$, $y_{IO}$). If the direct distance between T and IO is $d_{T\_IO}$, it can be inferred that M need to be moved by $d_{T\_IO}+1$ steps to reach its target IO. In this case, if there are more than $d_{T\_IO}+1$ escorts in the rectangle, it is enough to move M from its original position to its target IO. Otherwise, there are two solutions to get enough escorts to move M. On the one hand, the escorts outside the rectangle can be moved inside. On the other hand, the escorts in the rectangle can be reused after being used to move M. In either way, more moving steps are needed. This is the reason for introducing correction $r$ in the estimation of the elements in Matrix $\mathbf{D}_s$.

The number of required escorts $en_{IO,T}$ to move M to IO through T can be calculated as follows:

$$en_{IO,T} = e_{IO,T} - es_{IO,T} + 1 = |x_{IO} - x_T| + |y_{IO} - y_T| - es_{IO,T} + 1 \qquad (6)$$

where ($x_{IO}$, $y_{IO}$), ($x_M$, $y_M$) are the coordinates of IO and M, $es_{IO,T}$ is the number of escorts in the rectangle formed by IO and T, $e_{IO,T}$ is the distance between M and IO. If $en_{IO,T}$ is positive, correction $r$ should be introduced.

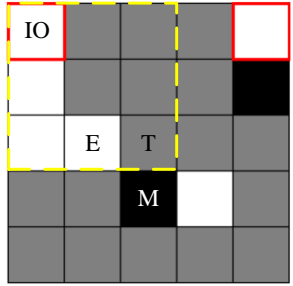

Fig. 6 An example where correction $r$ is needed.

An example is given in Fig. 6. In this case, the distance from T to IO is 4, while there are only 4 escorts in the rectangle. According to (6), one more escort is needed, and then correction $r$ is needed.

The detailed computation procedure of correction $r$ is given in Fig. 7. Two more example cases will be given to better illustrate the computation procedure.

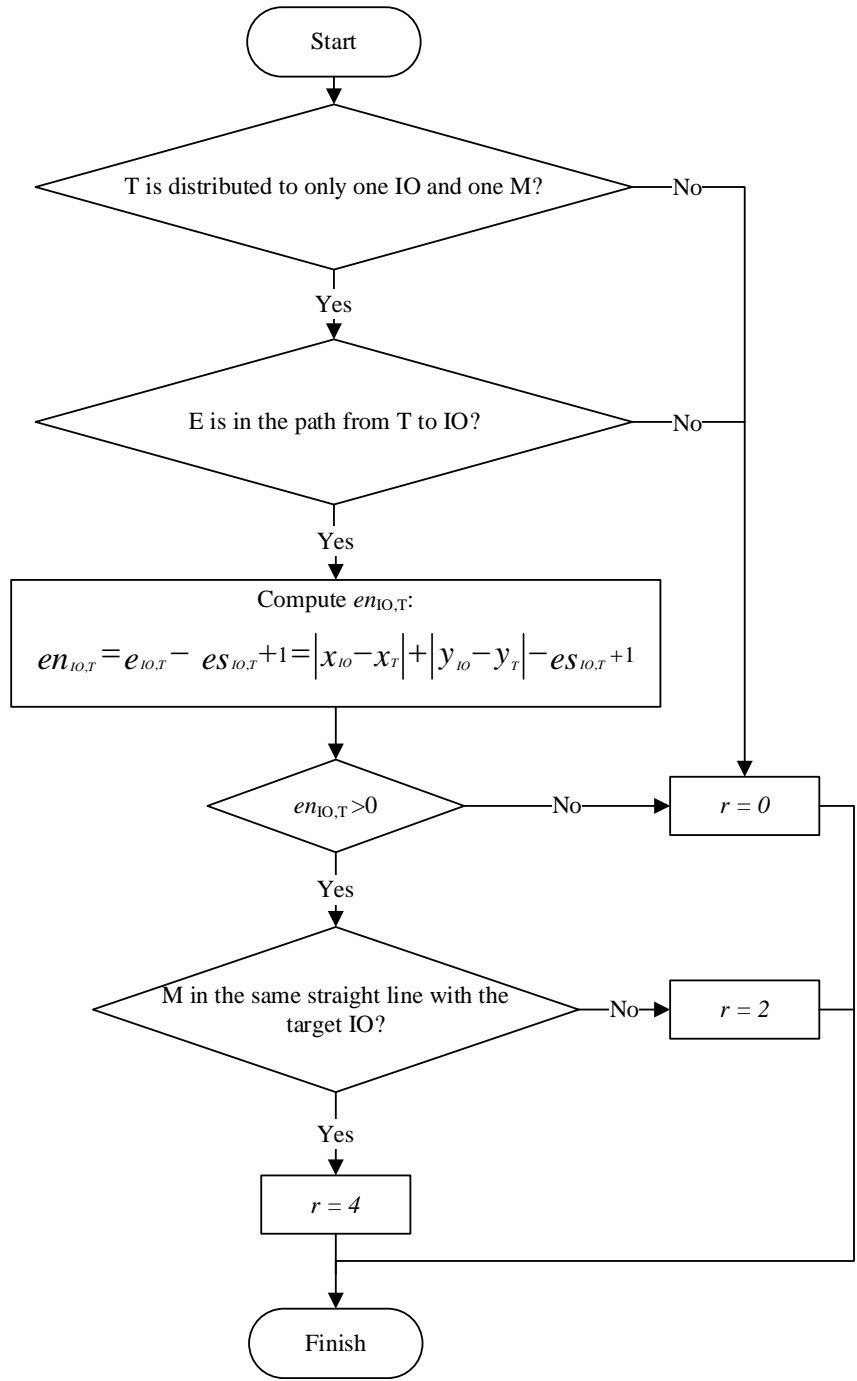

Fig. 7 Computation procedure of correction *r*.

(1) Example Case 1: *r*=4

The first example case is given in Fig. 8 and Fig. 9, in which two different routes are given. At the initial status, in order to move M towards the marked IO, two escorts, i.e. E and D, can be applied. For the escort target position T, the distance to E is 1 and the distance to D is 4 with correction *t* considered. Then, the choice without considering correction *r* is to apply E in order to move M. However, after moving

M one step towards the target IO, there is no escort which can be applied to move M. As a consequence, E has to be reused and the number of total moving steps is 7.

In contrast, if correction *r* is considered, the distance between E and T in the initial status is corrected as 5, according to the procedure in Fig. 7. Since the distance between D and T is 4, D will be selected to move M to T. The whole route is given in Fig. 9, in which the number of total moving steps is 6.

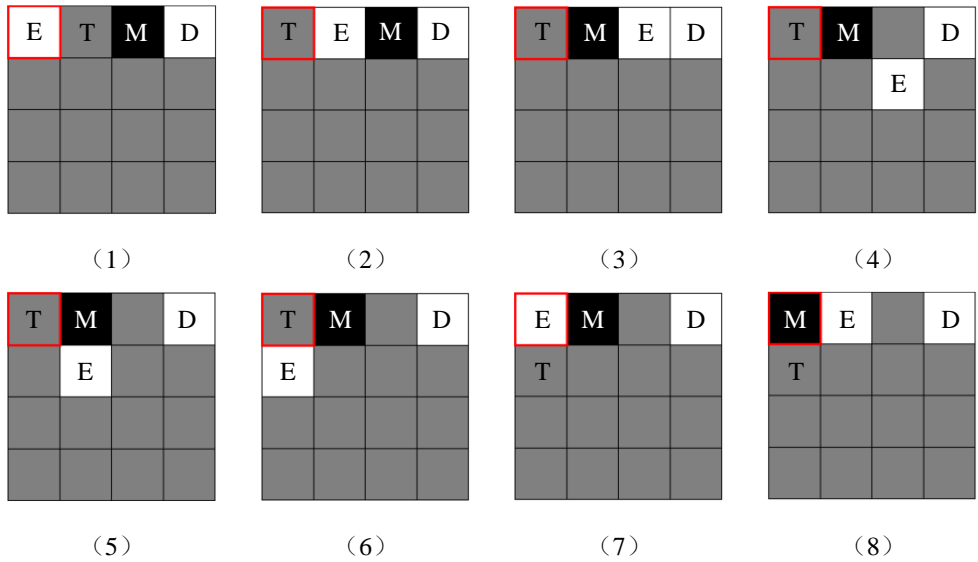

Fig. 8 Example Case 1, wrong route.

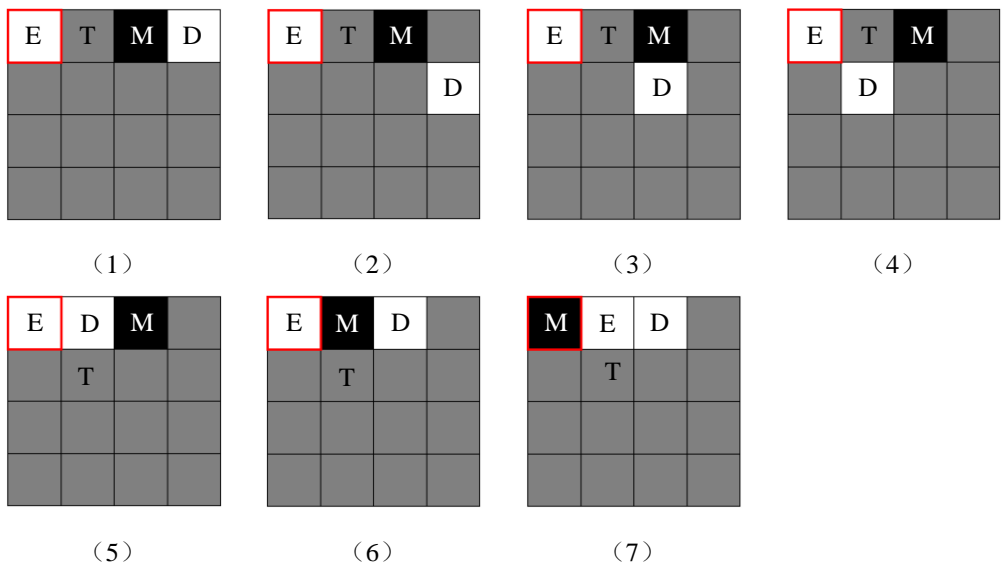

Fig. 9 Example Case 1, correct route.

(2) Example Case 2: *r*=2

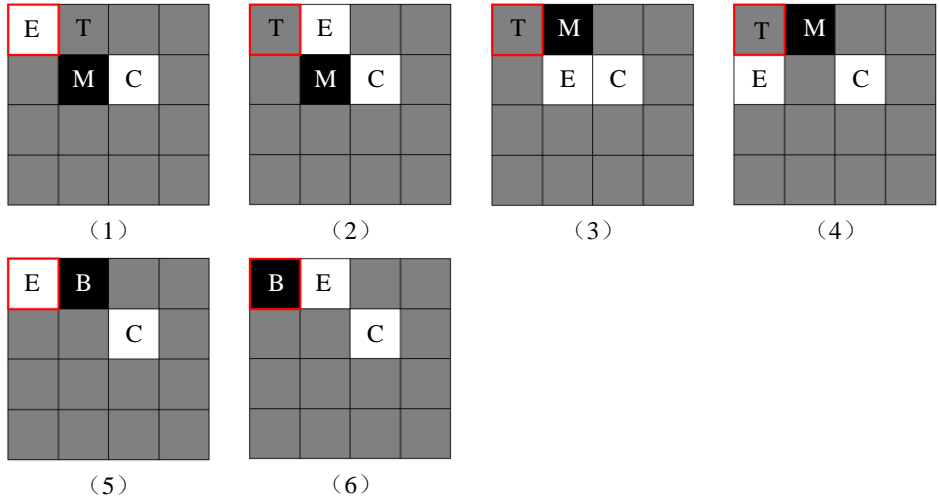

Fig. 10 Example Case 2, wrong route.

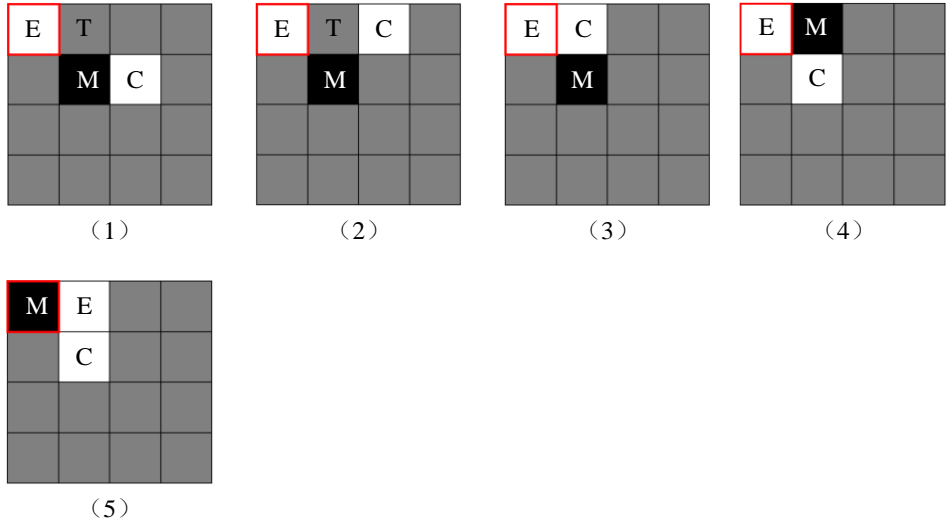

Fig. 11 Example Case 2, correct route.

The second case is an *r*=2 case. Similar to the first case, one moving step will be wasted if correction *r* is not considered. In Fig. 10, correction *r* is not considered and E is applied to move M to T. Afterwards, E has to be reused to further move M. In Fig. 11, correction *r* is considered, and C is applied to move M to T. In this route, one step is saved by avoiding the reuse of escorts.

To conclude, the introduction of correction *r* is to avoid the reuse of the escorts and to save the total moving steps. The function of correction *r* will be clearer in the following section, in which how to make the decisions according to $\mathbf{D}_s$ is illustrated.

### 4.4.3. Computation Procedures of Distance Matrix $\mathbf{D}_s$

It is time consuming to compute all the elements in $\mathbf{D}_s$. Since only the minimal element of $\mathbf{D}_s$ is useful in the decision making procedures, a vector $\mathbf{Q}$ is defined as follows, which is formed by the minimal elements of all the columns in $\mathbf{D}_s$:

$$\mathbf{Q} = [q_1, q_2, \ldots, q_m], \text{ where } q_j = \min(c_{i,j}), i=1, 2, \ldots, n.$$

Then, $\min(\mathbf{D}_s) = \min(\mathbf{Q})$

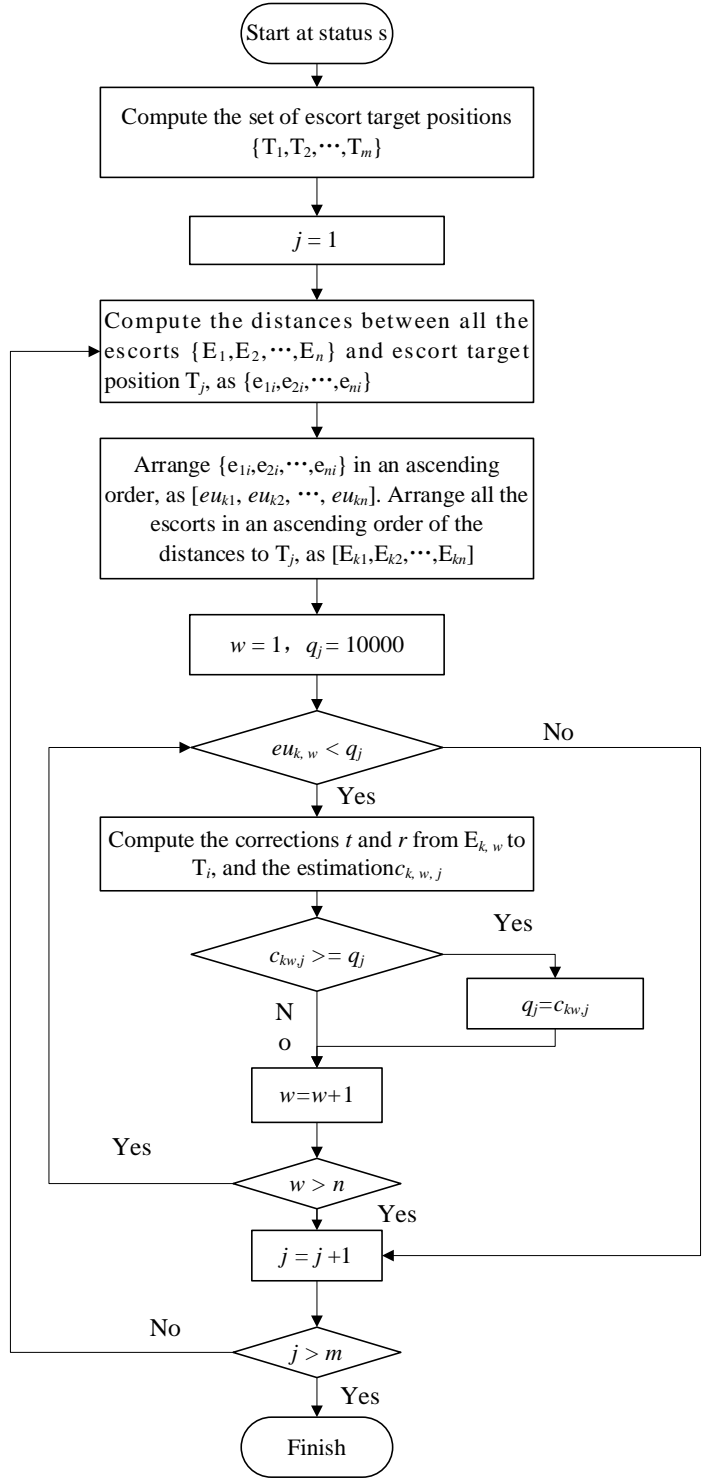

Fig. 12 Computation procedure of vector **Q**.

Afterwards, the computation procedures of vector **Q** are given in Fig. 12. The basic idea is that $q_j$ is the minimal distance from one escort target position to all the escorts. Then, for one escort target position, only the minimal distance to all the escorts, instead of all the distance to all the escorts, is needed to be computed. In this way, min($\mathbf{D}_s$) can be obtained, which is an important basis for decision making.

*4.5. Procedures of the Heuristic Algorithm*

As introduced previously, the heuristic algorithm is a solution to a MDP problem. Therefore, what is required from the heuristic algorithm is just the one step decision for a given status. In this section, the procedures of how to evaluate a status according to the indexes and make the one step decision is introduced.

### 4.5.1. Computation of Status Evaluation Indexes

To begin with, the current status should be analyzed and evaluated, after which the indexes can be computed as the basis for decision making. Three status evaluation indexes, i.e. the minimal total distance from the target items to the IOs $d_{s\_\min}$, the total number of required escorts $et_{s\_\min}$, and the minimal element of the distance matrix $\mathbf{D}_s$, have all been previously introduced. These indexes can be computed in the following 4 steps.

**Step 1**: Optimal distribution of target IOs. The details have been introduced in Section 3.1. The basic rule is to minimize the total distance from the target items to the IOs, afterwards $d_{s\_\min}$ can be obtained.

**Step 2**: Determination of escort target positions. The details have been introduced in Section 3.2.

**Step 3**: Computation of the total number of required escorts. The details have been introduced in Section 3.3. Afterwards, $et_{s\_\min}$ can be computed.

**Step 4**: Computation of the minimal element of the distance matrix $\mathbf{D}_s$. The details have been given in Section 3.4 and min($\mathbf{D}_s$) can be computed.

In this way, all the three indexes for one status are computed, which are the basis for further decision making procedure.

### 4.5.2. Adjacent Statuses Generation and Reward Computation

After evaluating one status $s$, its adjacent statuses set $\{s_i'\}$, $i=1, 2, …, n_s$, can be generated by moving one escort from $s$, where $n_s$ is the number of adjacent statuses of $s$. Then, all the elements in $\{s_i'\}$ can be evaluated in the same way as $s$, after which $d_{si'\_\min}$, $et_{si'\_\min}$ and $\mathbf{D}_{si'}$ for $i=1, 2, …, n_s$ can be obtained.

The next step is to compare the evaluation indexes between $s$ and each element in $\{s_i'\}$. Afterwards, $\{R_i\}$, $i=1, 2, …, n_s$, can be computed, where $R_i$ is the "reward" of transiting from $s$ to $s_i'$. The $s_i'$ with higher $R_i$ is better preferred to be the next status of $s$.

Among all the movements, the ones which can directly moving the target items to be nearer to the IOs are the most preferred. Therefore, the rewards of such movements are set to be the highest, i.e. 100. In other words, if $d_{si'\_\min} < d_{s\_\min}$, $R_i=100$. In contrast, $R_i=-1$ if $d_{si'\_\min} > d_{s\_\min}$, which means such movements should be avoided.

If $d_{s\_\min}$ cannot be directly reduced, it can be inferred that no escort is currently at the escort target positions. In such situations, the movements which can move the escorts towards the escort target positions should be preferred. Then, the movements which can reduce $et_{s\_\min}$ are the second preferred, and the movements which can reduce $\min(\mathbf{D}_s)$ are the third preferred. The rewards of these two kinds of movements are set as 50 and 10, respectively. In contrast, the movements which increase $et_{s\_\min}$ and $\min(\mathbf{D}_s)$ should be avoided, the rewards of which are set as -1. Afterwards, if none of $d_{s\_\min}$, $et_{s\_\min}$ and $\min(\mathbf{D}_s)$ are changed after one movement, the reward is set as 0.

To conclude, the rewards of different types of movements are given in TABLE II.

TABLE II Rewards of Different Types of Movements

| Type | Total distance $d_{s\_\min}$ | The number of required escorts $et_{s\_\min}$ | Minimal elements in distance matrix $\min(\mathbf{D}_s)$ | Reward |
|---|---|---|---|---|
| 1 | $d_{si'\_\min} < d_{s\_\min}$ | / | / | 100 |
| 2 | $d_{si'\_\min} > d_{s\_\min}$ | / | / | -1 |
| 3 | $d_{si'\_\min} = d_{s\_\min}$ | $et_{si'\_\min} < et_{s\_\min}$ | / | 50 |
| 4 | $d_{si'\_\min} = d_{s\_\min}$ | $et_{si'\_\min} > et_{s\_\min}$ | / | -1 |
| 5 | $d_{si'\_\min} = d_{s\_\min}$ | $et_{si'\_\min} = et_{s\_\min}$ | $\min(\mathbf{D}_{si'}) < \min(\mathbf{D}_s)$ | 10 |
| 6 | $d_{si'\_\min} = d_{s\_\min}$ | $et_{si'\_\min} = et_{s\_\min}$ | $\min(\mathbf{D}_{si'}) > \min(\mathbf{D}_s)$ | -1 |
| 7 | $d_{si'\_\min} = d_{s\_\min}$ | $et_{si'\_\min} = et_{s\_\min}$ | $\min(\mathbf{D}_{si'}) = \min(\mathbf{D}_s)$ | 0 |

4.5.3. Decision Making Procedure

The decision making procedure of the proposed heuristic algorithm from a given status $s$ is given in …. Firstly, the evaluation indexes $d_{s\_\min}$, $et_{s\_\min}$ and $\min(\mathbf{D}_s)$ are computed. Secondly, the adjacent statuses of $s$ $\{s_i'\}$, are generated and evaluated. Thirdly, each element in $\{s_i'\}$ is compared with $s$ and then $R_i$ can be obtained. Finally, the decision is made as $\operatorname{argmax}(R_i)$, which means the status $s_i'$ with the largest $R_i$ is chosen as the next status to $s$. In this way, this procedure is continued until all the target elements arrive at the IOs.

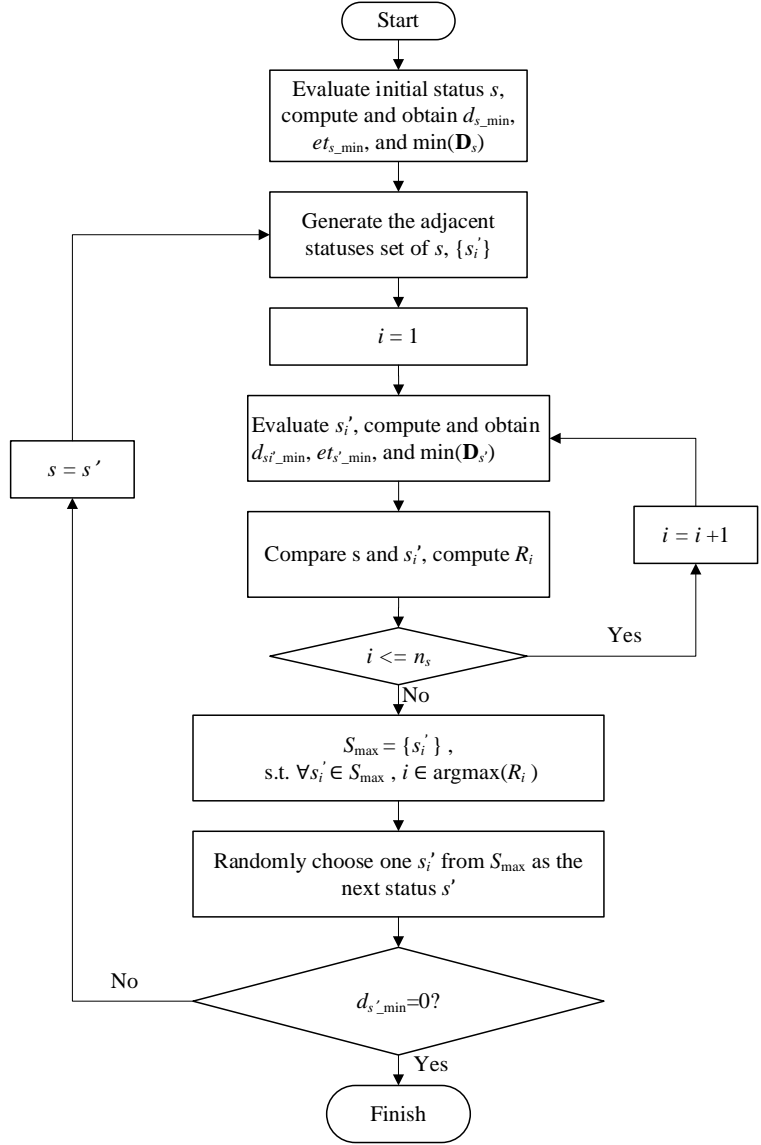

Fig. 13 Decision making procedure of the heuristic algorithm.

## 5. Numerical Results

### 5.1. Simulation System Introduction

Since there is no PBSS simulation platform that can be directly used so far, in order to analyse the experimental process and results more intuitively, this paper builds a PBSS simulation platform as shown in Fig. 14 (a) based on Python 3.7. The functions of the simulation platform can be summarized as follows:

**1. Embed a heuristic algorithm:** The PBSS simulation platform embeds the heuristic algorithm proposed in this article. Users can choose the solution algorithm in the interface.

**2. Set map information:** As shown in Fig. 14 (b), in the visual interface, the user can set the map information used by the PBSS, including: map size, number of escorts, and IO locations.

**3. Mouse click to set an initial state:** The user can directly set the type of the block in the map by clicking the left and right buttons of the mouse, including: target items, other items and escorts. In this way, the user can perform the initial state according to the needs.

**4. Randomly generate an initial state:** In order to facilitate numerical experiments, the system can randomly generate the initial state of a map with map information.

**5. Looking back at the routing decision process:** To look back at the results of the heuristic algorithm intuitively, users can click the "Previous" and "Next" buttons, so that the entire routing decision process can be shown.

The numerical experiments in this article are all based on an Intel CORE i5 6200 8G RAM computer.

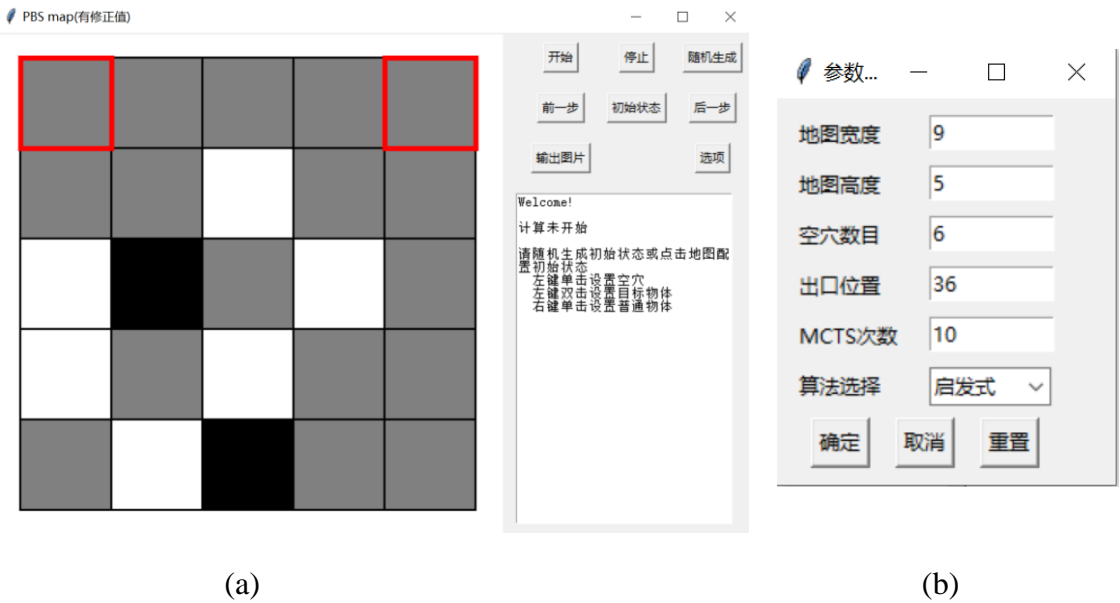

(a) (b)

Fig. 14 Interface of PBSS simulation platform.

*5.2. Example Cases*

5.2.1. An Example Case with Single Picking Items

The routing decisions of a one-item-one-IO PBSS computing by the heuristic algorithm with correction $r$ and correction $t$ are shown in TABLE III. There is only one IO (0, 0) in the PBSS and thus the target IO of the target item is always (0, 0). The result of the routing solution and the decision reason and reward for each step are also shown in TABLE III.

For example, as shown in Fig. 15, in the initial state (step 0), the minimal moving step estimation $c$ is from escort (3, 0) to escort target (3, 3), thus the decision reason of the first step (step1) shown in this problem is the minimal elements in distance matrix $\min(\mathbf{D}_s)$. After moving the escort from (3, 0) to (3, 1), $c$ is changed to 2 from 3 and the reward is 10. The second moving decision is to move the escort from (3, 1) to (2, 2), with the decision reason of the number of required escorts $et_{s\_min}$. After moving the escort from (3, 1) to (2, 2), The number of required escorts $et_{s\_min}$ is changed from 2 to 1. The fourth moving decision is to move the escort from (2, 2) to (2, 3), with the decision reason of the total distance $d_{s\_min}$. After moving the escort from (2, 2) to (2, 3), the total distance $d_{s\_min}$ is changed from 5 to 4.

TABLE III The Decision Reason and Reward of a One-Item-One-IO PBSS Routing Problem

| Step | Escort from | Escort to | Reason | Value before move | Value after move | Reward |
|---|---|---|---|---|---|---|
| 1 | (3, 0) | (3, 1) | Minimal elements in distance matrix min($\mathbf{D}_s$) | 3 | 2 | 10 |
| c | (3, 1) | (2, 2) | Number of required escorts $et_{s\_min}$ | 2 | 1 | 50 |
| 3 | (2, 1) | (2, 2) | Minimal elements in distance matrix min($\mathbf{D}_s$) | 3 | 2 | 10 |
| 4 | (2, 2) | (2, 3) | Total distance $d_{s\_min}$ | 5 | 4 | 100 |
| 5 | (2, 3) | (1, 3) | Minimal elements in distance matrix min($\mathbf{D}_s$) | 2 | 1 | 10 |
| 6 | (1, 3) | (1, 2) | Number of required escorts $et_{s\_min}$ | 1 | 0 | 50 |
| 7 | (1, 2) | (2, 2) | Total distance $d_{s\_min}$ | 4 | 3 | 100 |
| 8 | (2, 2) | (2, 1) | Minimal elements in distance matrix min($\mathbf{D}_s$) | 2 | 1 | 10 |
| 9 | (2, 1) | (1, 1) | Number of required escorts $et_{s\_min}$ | 1 | 0 | 50 |
| 10 | (1, 1) | (1, 2) | Minimal elements in distance matrix min($\mathbf{D}_s$) | 3 | 2 | 100 |
| 11 | (0, 1) | (1, 1) | Minimal elements in distance matrix min($\mathbf{D}_s$) | 2 | 1 | 100 |
| 12 | (0, 0) | (0, 1) | Minimal elements in distance matrix min($\mathbf{D}_s$) | 1 | 0 | 100 |

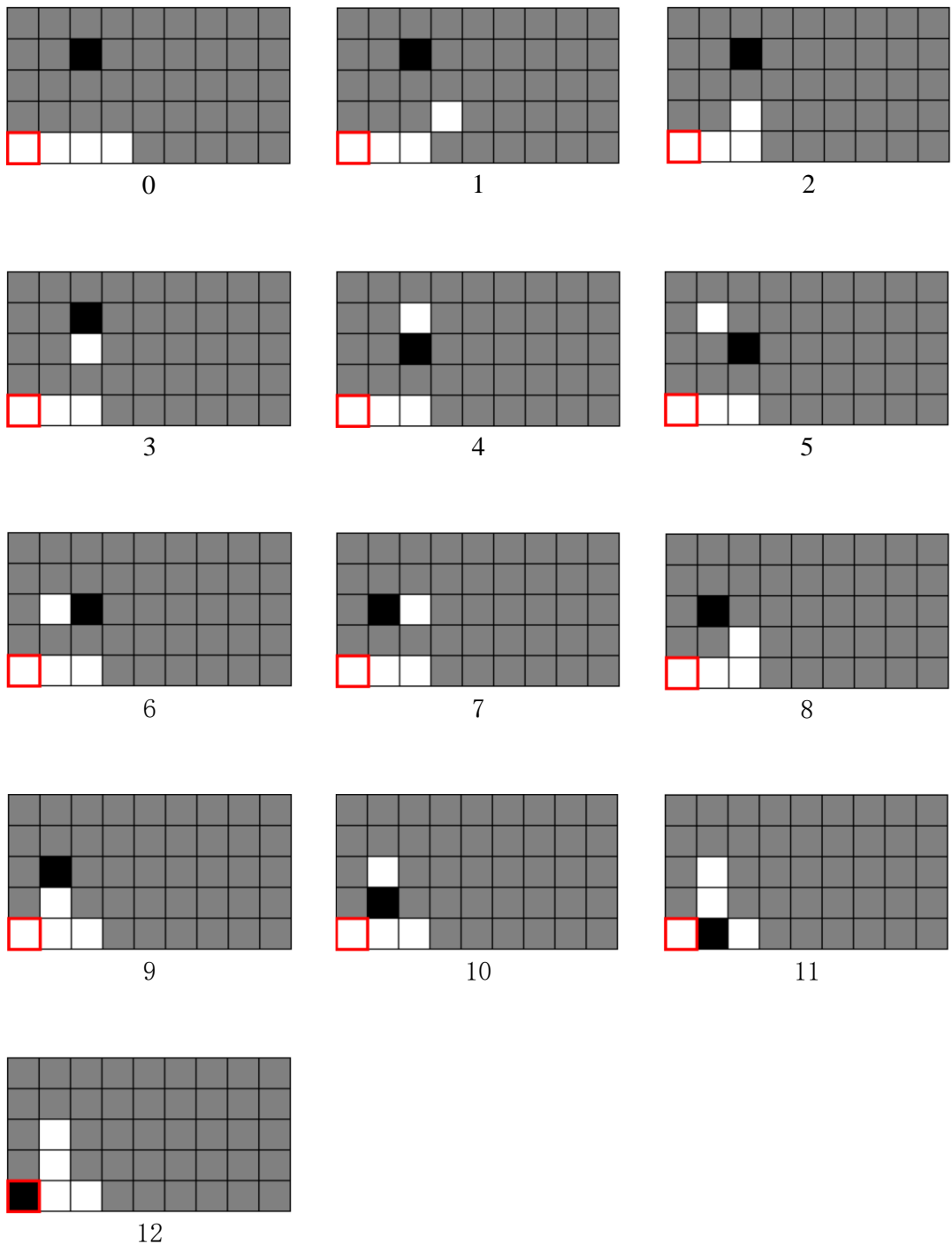

Fig. 15 The computing results of a One-Item-One-IO PBSS Routing Problem with heuristic algorithm.

5.2.2. An Example Case with Multiple Picking Items

The routing decisions of a multiple-item-multiple-IO PBSS computing by the heuristic algorithm with correction *r* and correction *t* are shown in TABLE IV. There are two IOs (0, 3), (3, 3) in the PBSS. And thus there are 2 arrangements of the target IOs for the target items, that is, the upper and lower target items correspond to the left and right IOs respectively. Both of the total distance between the target items and the target IOs under these the two allocations are 6. The result of the routing solution and the decision reason and reward for each step are also shown in TABLE IV.

For example, as shown in Fig. 16, in the initial state (step 0), the minimum moving step estimation $c$ is from escort (0, 2) to escort target (1, 2), thus the decision reason of the first step (step1) shown in this problem is the total distance $d_{s\_\min}$. After moving the escort from (0, 2) to (1, 2), the number of required escorts $et_{s\_\min}$ is changed to 6 from 5 and the reward is 100. The third moving decision is to move the escort from (1, 1) to (2, 1), with the decision reason of the minimal element in distance matrix $\min(\mathbf{D}_s)$. After moving the escort from (1, 1) to (2, 1), the minimal element in distance matrix is changed from 2 to 1. The fourth moving decision is to move the escort from (2, 1) to (2, 2), with the decision reason of the number of required escorts $et_{s\_\min}$. After moving the escort from (2, 1) to (2, 2), the number of required escorts $et_{s\_\min}$ is changed from 3 to 2.

TABLE IV The Decision Reason and Reward of a Multiple-Item-Multiple-IO PBSS Routing Problem

| Step | Escort from | Escort to | Reason | Value before move | Value after move | Reward |
| --- | --- | --- | --- | --- | --- | --- |
| 1 | (0, 2) | (1, 2) | Total distance $d_{s\_\min}$ | 6 | 5 | 100 |
| 2 | (1, 2) | (1, 1) | Total distance $d_{s\_\min}$ | 5 | 4 | 100 |
| 3 | (1, 1) | (2, 1) | Minimal elements in distance matrix $\min(\mathbf{D}_s)$ | 2 | 1 | 10 |
| 4 | (2, 1) | (2, 2) | Number of required escorts $et_{s\_\min}$ | 3 | 2 | 50 |
| 5 | (2, 2) | (1, 2) | Total distance $d_{s\_\min}$ | 4 | 3 | 100 |
| 6 | (1, 2) | (1, 3) | Minimal elements in distance matrix $\min(\mathbf{D}_s)$ | 2 | 1 | 10 |
| 7 | (1, 3) | (2, 3) | Number of required escorts $et_{s\_\min}$ | 2 | 1 | 50 |
| 8 | (2, 3) | (2, 2) | Total distance $d_{s\_\min}$ | 3 | 2 | 100 |
| 9 | (3, 3) | (2, 3) | Total distance $d_{s\_\min}$ | 2 | 1 | 100 |
| 10 | (2, 3) | (1, 3) | Minimal elements in distance matrix $\min(\mathbf{D}_s)$ | 2 | 1 | 10 |
| 11 | (1, 3) | (0, 3) | Number of required escorts $et_{s\_\min}$ | 1 | 0 | 50 |
| 12 | (0, 3) | (0, 2) | Total distance $d_{s\_\min}$ | 1 | 0 | 100 |

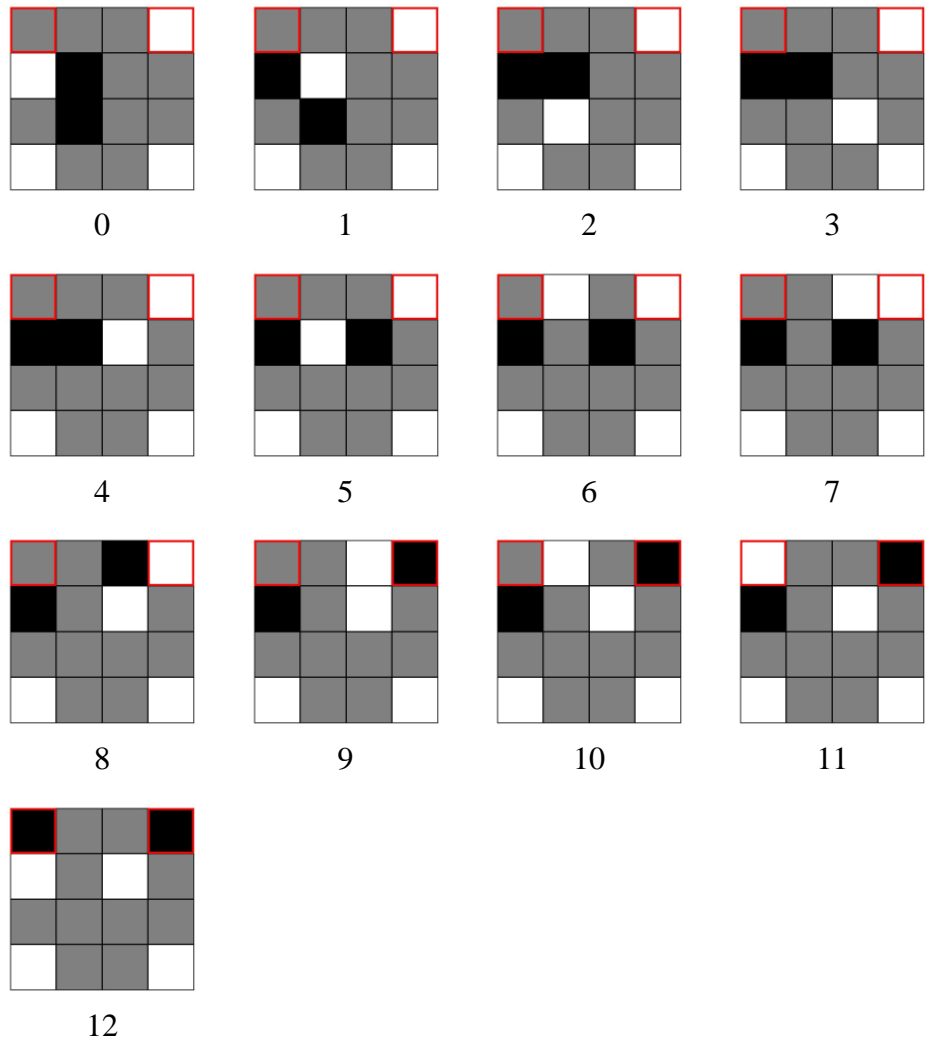

Fig. 16 The computing results of a Multiple-Item-Multiple-IO PBSS Routing Problem with heuristic algorithm

### 5.3. Results Analysis of Different Cases

#### 5.3.1. Single-Item Cases

To solve the problem of single target item and single IO, this paper compares the results of the calculation examples given by [1]. The algorithm proposed in [1], however, requires that the starting positions of all escorts are located in the lower left corner, arranged in a row, and the only IO is located in the lower left corner. Although the algorithm proposed in this article is not limited to this, in order to facilitate comparison, the same initial state settings as in [1]. will be used for experimental verification.

The PBSS map size of the experiment is 9×5, the number of escorts is 1-6, the number of target items is 1, and the position of the IO (0, 0). Taking the randomness of the algorithm into account, each example is run 3 times, and the statistical results of the experiments are the average of the three experimental results.

To measure the difference between the results by the proposed heuristic algorithm and those by the exact algorithm in [1], a metric named gap is defined as follows,

$$gap = \sum_{i=1}^{n} \Delta_i / n \qquad (7)$$

where: $\Delta_i$ is the relative difference in percentage between the number of steps by the proposed heuristic algorithm and that by the exact algorithm in [1] of the $i$th case, and $n$ is the total number of cases.

As shown in Fig. 17 the white blocks in the figure represent the starting positions of the escort, and the grey blocks represent the object items and other items. In each experiment, there is only one target item. The numbers on the grey blocks in the figure represent the number of steps required to move the target item to the IO from that position according to the algorithms. The white numbers represent the theoretical shortest number of steps which is also the result computing by the heuristic algorithm in this paper. The black numbers represent that the heuristic algorithm did not obtain the optimal solution and the upper number is the moving steps calculated by the algorithm while the number below is the theoretical shortest number of moving steps.

| 19 | 21 | 23 | 25 | 29 | 33 | 39 | 45 | 51 |
| 13 | 15 | 17 | 21 | 25 | 31 | 37 | 43 | 49 |
| 7 | 9 | 13 | 17 | 23 | 29 | 35 | 41 | 47 |
| 1 | 5 | 9 | 15 | 21 | 27 | 33 | 39 | 45 |
| | 1 | 7 | 13 | 19 | 25 | 31 | 37 | 43 |

One escort, average computation time
0.364s, Gap = 0

| 16 | 16 | 18 | 22 | 26 | 30 | 34 | 40 | 46 |
| 10 | 10 | 14 | 18 | 22 | 26 | 32 | 38 | 44 |
| 4 | 6 | 10 | 14 | 18 | 24 | 30 | 36 | 42 |
| 1 | 2 | 6 | 10 | 16 | 22 | 28 | 34 | 40 |
| | | 2 | 8 | 14 | 20 | 26 | 32 | 38 |

Two escorts, average computation time
0.356s, Gap = 0

| 15 | 15 | 17 (16) | 19 | 23 | 27 | 31 | 36 (35) | 41 |
| 9 | 11 (10) | 11 | 15 | 19 | 23 | 27 | 33 | 39 |
| 4 | 5 | 7 | 11 | 15 | 19 | 25 | 31 | 37 |
| 1 | 2 | 3 | 7 | 11 | 17 | 23 | 29 | 35 |
| | | | 3 | 9 | 15 | 21 | 27 | 33 |

Three escorts, average computation time
0.465s, Gap = 0.45%

| 15 | 15 | 16 | 18 (17) | 20 | 24 | 28 | 32 | 38 (36) |
| 9 | 11 (10) | 12 (11) | 12 | 16 | 20 | 24 | 28 | 34 |
| 4 | 5 | 6 | 8 | 12 | 16 | 20 | 26 | 32 |
| 1 | 2 | 3 | 4 | 8 | 12 | 18 | 24 | 30 |
| | | | | 4 | 10 | 16 | 22 | 28 |

Four escorts, average computation time
0.345s, Gap = 0.74%

| 15 | 15 | 16 | 17 | 19 (18) | 21 | 25 | 29 | 33 |
| 9 | 11 (10) | 12 (11) | 13 (12) | 13 | 17 | 21 | 25 | 29 |
| 4 | 5 | 6 | 7 | 9 | 13 | 17 | 21 | 27 |
| 1 | 2 | 3 | 4 | 5 | 9 | 13 | 19 | 25 |
| | | | | | 5 | 11 | 17 | 23 |

Five escorts, average computation time
0.343s, Gap = 0.82%

| 15 | 15 | 16 | 17 | 19 (18) | 20 (19) | 22 | 26 | 30 |
| 9 | 11 (10) | 12 (11) | 13 (12) | 14 (13) | 14 | 18 | 22 | 26 |
| 4 | 5 | 6 | 7 | 8 | 10 | 14 | 18 | 22 |
| 1 | 2 | 3 | 4 | 5 | 6 | 10 | 14 | 20 |
| | | | | | | 6 | 12 | 18 |

Six escorts, average computation time
0.465s, Gap = 1.18%

Fig. 17 Experimental Results of PBSS Routing Problem with the Method of Heuristic Algorithm

When solving the PBSS routing problem, the heuristic algorithm proposed in this paper can obtain the theoretical optimal solution for 94.08% cases. The average gap between the heuristic solution and the exact solution is only 0.53%. This shows that the heuristic algorithm proposed in this paper can obtain a feasible solution close to the exact solution in a short time, and proves the validity and correctness of the algorithm.

### 5.3.2. Results Analysis of Multiple-Item Cases

As there is no method to solve the multi-item-multi-IO PBSS routing problem in the previous literature, the numerical experiments of multiple-item cases lack accurate solutions to the PBSS routing problems. The heuristic algorithm proposed in this paper can solve the multi-item-multi-IO PBSS routing problem with a map size of 9*9.

The results are shown in TABLE V and the details of the initial states of the PBSS in the routing problems are shown in the appendix.

TABLE V Experiment Results of Multi-item-multi-IO PBSS Routing Problem

| Case | Experiment Results (number of steps) | Computing Time(s) |
|---|---|---|
| 9-1 | 49 | 2.67 |
| 9-2 | 38 | 2.91 |
| 9-3 | 41 | 3.49 |
| 9-4 | 23 | 1.14 |
| 9-5 | 50 | 4.79 |
| 9-6 | 49 | 7.30 |
| 9-7 | 44 | 9.94 |
| 9-8 | 32 | 5.39 |
| 9-9 | 28 | 4.56 |
| 9-10 | 38 | 5.72 |
| 9-11 | 54 | 10.00 |
| 9-12 | 40 | 7.94 |
| 9-13 | 39 | 14.06 |
| 9-14 | 37 | 12.76 |
| 9-15 | 23 | 5.76 |
| Average | 39 | 6.56 |

6. **Conclusion**

In order to solve the problem of limited storage space and high rental cost faced by logistics enterprises, this paper studies the route programming problem of PBSS, defines the description of the PBSS load retrieval problem, and proposes an MDP based heuristic algorithm for PBSS load retrieval route programming. This paper innovatively solves the problem with multiple target items in a PBSS, with the cooperation and coordination of different escorts considered. The case study results have validated the ability of the proposed algorithm to efficiently solve PBSS load retrieval route programming problems with multiple target items.

This work can provide a fast solution to a complicated PBSS load retrieval route programming problem. Its high computation efficiency is favoured by reinforcement learning algorithms such as Monte-Carlo tree search algorithm and deep reinforcement learning. In future work, the proposed algorithm in this paper can be regarded as the basis of applying reinforcement learning algorithms to solve PBSS load retrieval route programming problems, through which the solution of route programming can be further improved.

8. **Appendix**

TABLE VI PBSS cases with multiple target items in a 9×9 map

| Case | Number of escorts | Number of target items | Location of IO | Initial status |
|---|---|---|---|---|
| 9-1 | 9 | 3 | (0, 0); (0, 8); (8, 8) | 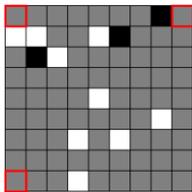 |
| 9-2 | 9 | 3 | (0, 0); (0, 8); (8, 8) | 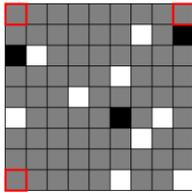 |
| 9-3 | 9 | 3 | (4, 0); (8, 8); (0, 8) | 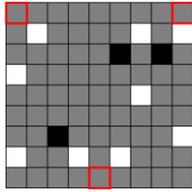 |
| 9-4 | 9 | 3 | (4, 0); (8, 8); (0, 8) | 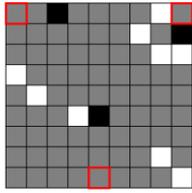 |
| 9-5 | 9 | 3 | (0, 8); (8, 8); (0, 8) | 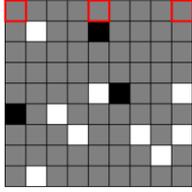 |
| 9-6 | 15 | 4 | (0, 0); (0, 8); (8, 8); (0, 8) | 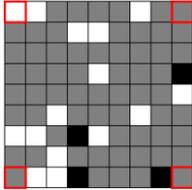 |
| 9-7 | 15 | 4 | (0, 0); (0, 8); (8, 8); (0, 8) | 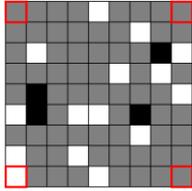 |
| 9-8 | 15 | 4 | (4, 0); (8, 4); (4, 8); (0, 4) | 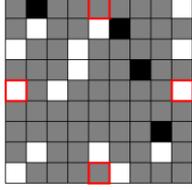 |
| 9-9 | 15 | 4 | (4, 0); (8, 4); (4, 8); (0, 4) | 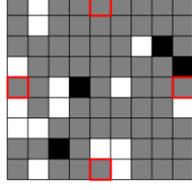 |
| 9-10 | 15 | 4 | (0, 4); (0, 8); (8, 8); (8, 4) | 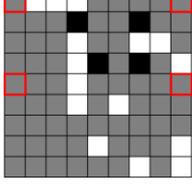 |
| 9-11 | 20 | 4 | (0, 0); (8, 0); (8, 8); (0, 8); | 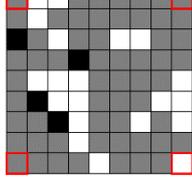 |
| 9-12 | 20 | 4 | (0, 0); (8, 0); (8, 8); (0, 8); | 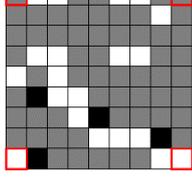 |
| 9-13 | 20 | 4 | (4, 0); (8, 4); (4, 8); (0, 4); | 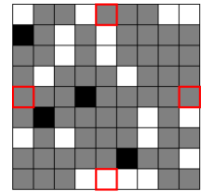 |
| 9-14 | 20 | 4 | (4, 0); (8, 4); (4, 8); (0, 4); | 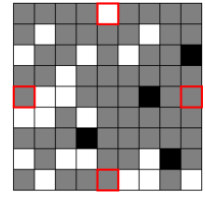 |
| 9-15 | 20 | 4 | (4, 0); (8, 4); (4, 8); (0, 4); | 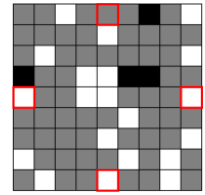 |